\documentclass[onecolumn, a4size, 11pt]{IEEEtran}
\usepackage{amsmath}
\usepackage{amssymb}
\usepackage{amsfonts}
\usepackage{graphicx}
\usepackage{epsfig}
\usepackage{subfigure}
\usepackage{psfrag}
\usepackage{xcolor}

\title{Wireless Information Transfer with Opportunistic Energy
Harvesting \footnote{This paper has been presented in part at IEEE International Symposium on Information Theory (ISIT),
Cambridge, MA, USA, July 1-6, 2012.}\footnote {L. Liu is with the Department of Electrical
and Computer Engineering, National University of Singapore
(e-mail:liu\_liang@nus.edu.sg).}\footnote{R. Zhang is with the
Department of Electrical and Computer Engineering, National
University of Singapore (e-mail:elezhang@nus.edu.sg). He is also
with the Institute for Infocomm Research, A*STAR, Singapore.}
\footnote{K. C. Chua is with the Department of Electrical and
Computer Engineering, National University of Singapore
(e-mail:eleckc@nus.edu.sg).}}

\author{Liang Liu, Rui Zhang, and Kee-Chaing Chua}

\begin{document}
\maketitle \thispagestyle{empty} \vspace{-0.3in}

\begin{abstract}
Energy harvesting is a promising solution to prolong the operation
of energy-constrained wireless networks. In particular, scavenging
energy from ambient radio signals, namely \emph{wireless energy
harvesting} (WEH), has recently drawn significant attention. In this
paper, we consider a point-to-point wireless link over the narrowband
flat-fading channel subject to time-varying co-channel
interference. It is assumed that the receiver has no fixed power
supplies and thus needs to replenish energy opportunistically via WEH from the
unintended interference and/or the intended signal sent by the
transmitter. We further assume a single-antenna receiver that can
only decode information or harvest energy at any time due to
the practical circuit limitation. Therefore, it is important to
investigate when the receiver should switch between the two modes of
information decoding (ID) and energy harvesting (EH), based on the
instantaneous channel and interference condition. In this paper, we
derive the optimal mode switching rule at the receiver to achieve
various trade-offs between wireless information transfer and energy
harvesting. Specifically, we determine the minimum transmission
outage probability for delay-limited information transfer and the
maximum ergodic capacity for no-delay-limited information transfer
{\it versus} the maximum average energy harvested at the receiver, which
are characterized by the boundary of so-called ``outage-energy''
region and ``rate-energy'' region, respectively. Moreover, for the
case when the channel state information (CSI) is known at the
transmitter, we investigate the joint optimization of transmit power
control, information and energy transfer scheduling, and the
receiver's mode switching. The effects of circuit energy consumption at the receiver on the achievable rate-energy trade-offs are also characterized. Our results provide useful guidelines for
the efficient design of emerging wireless communication systems
powered by opportunistic WEH.

\end{abstract}


\setlength{\baselineskip}{1.0\baselineskip}
\newtheorem{definition}{\underline{Definition}}[section]
\newtheorem{fact}{Fact}
\newtheorem{assumption}{Assumption}
\newtheorem{theorem}{\underline{Theorem}}[section]
\newtheorem{lemma}{\underline{Lemma}}[section]
\newtheorem{corollary}{\underline{Corollary}}[section]
\newtheorem{proposition}{\underline{Proposition}}[section]
\newtheorem{example}{\underline{Example}}[section]
\newtheorem{remark}{\underline{Remark}}[section]
\newtheorem{algorithm}{\underline{Algorithm}}[section]
\newcommand{\mv}[1]{\mbox{\boldmath{$ #1 $}}}

\section{introduction}
In conventional energy-constrained wireless networks such as sensor
networks, the lifetime of the network is an important performance
indicator since sensors are usually equipped with fixed energy supplies,
e.g., batteries, which are of limited operation time. Recently,
energy harvesting has become an appealing solution to prolong the
lifetime of wireless networks. Unlike battery-powered networks,
energy-harvesting wireless networks potentially have an unlimited
energy supply from the environment. Consequently, the research of wireless networks
powered by renewable energy has recently drawn a great deal of
attention (see e.g. \cite{Kulkarni11} and references therein).

In addition to other commonly used energy sources such as solar and
wind, ambient radio signals can be a viable new source for wireless
energy harvesting (WEH). Since radio signals carry information as
well as energy at the same time, an interesting new research
direction, namely ``simultaneous wireless information and power
transfer'', has recently been pursued
\cite{Varshney08}-\cite{Rui11}. The above prior works have studied
the fundamental performance limits of wireless information and
energy transfer systems under different channel setups, where the
receiver is assumed to be able to decode the information and harvest
the energy from the same signal, which may not be realizable yet due
to practical circuit limitations \cite{Rui11}. Consequently, a so-called ``time switching'' scheme, where the receiver switches over time between decoding information and harvesting energy, was proposed in \cite{Rui11} and \cite{Varshney} as a practical design. In this paper, we investigate further the time-switching scheme for a point-to-point single-antenna flat-fading channel subject to
time-varying co-channel interference, as shown in Fig. \ref{fig1}. Our motivations for investigating time switching are as follows. Firstly, with time switching, off-the-shelf commercially available circuits that are separately designed for information decoding and energy harvesting can be used, thus reducing the receiver's complexity as compared to other existing designs, e.g., ``power splitting" \cite{Rui11} and ``integrated receiver" \cite{Rui12}. Secondly, time switching judiciously exploits the facts that (1) information and energy receivers in practice operate with very different power sensitivity (e.g., -10dBm for energy receivers versus -60dBm for information receivers); and (2) wireless transmissions typically experience time-varying channels (e.g., due to shadowing and fading) and/or interferences (e.g., in a spectrum sharing environment), which fluctuate in very large power ranges (e.g., tens of dBs). Therefore, a time-switching receiver can utilize both the energy/information receiver power sensitivity difference and channel/interference power dynamics to optimize its switching operation. For example, the receiver can be switched to harvest energy when the channel (or interference) is strong, or decode information  when the channel (or interference) is relatively weaker.

In this paper, we assume that the transmitter has a fixed power supply (e.g.,
battery), whereas the receiver has no fixed power supplies and thus
needs to replenish energy via WEH from the received interference
and/or signal sent by the transmitter. We consider an
\emph{opportunistic} WEH at the single-antenna receiver, i.e., the
receiver can only decode information or harvest energy at any given
time, but not both. As a result, the receiver needs to decide when
to switch between an information decoding (ID) mode and an energy
harvesting (EH) mode, based on the instantaneous channel gain and
interference power, which are assumed to be perfectly known at the
receiver. In this paper, we derive the optimal mode switching rule
at the receiver to achieve various trade-offs between the minimum
transmission outage probability (if the information transmission is
delay-limited) or the maximum ergodic capacity (if the information
transmission is not delay-limited) in ID mode versus the maximum
average harvested energy in EH mode, which are characterized by the
boundary of the so-called ``outage-energy (O-E)'' region and
``rate-energy (R-E)'' region, respectively. Moreover, for the case
when the channel state information (CSI) is known at both the
transmitter and the receiver, we examine the optimal design of
transmit power control and scheduling for information and energy
transfer jointly with the receiver's mode switching, to achieve
different boundary pairs of the O-E region or R-E region. One important property of the proposed optimal resource allocation scheme is that the received signals with large power should be switched to the EH mode rather than ID mode, which is consistent with the fact that the energy receiver in general has a poorer sensitivity (larger received power) than the information receiver.

It is worth noting that from a traditional viewpoint, interference
is an undesired phenomenon in wireless communication since it
jeopardizes the wireless channel capacity if not being decoded and
subtracted completely. In the literature, fundamental approaches have been applied to deal with the interference in wireless information transfer, e.g., decoding the interference when it is strong \cite{Han81} or treating the interference as noise when it is weak \cite{Shang, Annapureddy}.
Recently, another approach, namely ``interference alignment'', was
proposed \cite{Jafar08}, where interference signals are properly aligned in a certain subspace of the received signal at each receiver to achieve the maximum degrees of freedom (DoF) for the sum-rate. Different from the
above works, this paper provides a new approach to deal
with the interference by utilizing it as a new source for WEH.
However, the fundamental role of interference in emerging wireless
networks with simultaneous information and power transfer still
remains unknown and is thus worth further investigation.


It is also worth pointing out that recently, another line of
research on wireless communication with energy-harvesting nodes has
been pursued (see e.g. \cite{rui11}-\cite{Ozel11} and references
therein). These works have addressed energy management policies at
the transmitter side subject to intermittent and random harvested
energy, which are thus different from our work that mainly
addresses opportunistic wireless energy harvesting at the receiver side.

The rest of this paper is organized as follows. Section
\ref{sec:system model and problem formulation} presents the system
model and illustrates the encoding and decoding schemes for wireless
information transfer with opportunistic energy harvesting. Section
\ref{sec:Information Transfer verse Energy Harvesting Trade-offs in
Fading Channels} defines the O-E and R-E regions and formulates the
problems to characterize their boundaries. Sections \ref{sec:Outage
Probability verse Averaged Harvested Energy} and \ref{sec:Ergodic
Capacity verse Averaged Harvested Energy} present the optimal mode
switching rules at the receiver, and power control and scheduling
polices for information and energy transfer at the transmitter (if
CSI is known) to achieve various O-E and R-E trade-offs,
respectively. Section \ref{sec:The Case When the Circuit Power Consumption is Considered for Both ID Mode and EH Mode} extends the optimal decision rule of the receiver to the case where the receiver energy consumption is taken into consideration. Section \ref{sec:simulation results} provides
numerical results to evaluate the performance of the proposed
schemes as compared against other heuristic schemes. Finally,
Section \ref{sec:concluding remarks} concludes the paper.

\section{System Model}\label{sec:system model and problem formulation}

\begin{figure}
\centering
 \epsfxsize=0.8\linewidth
    \includegraphics[width=12cm]{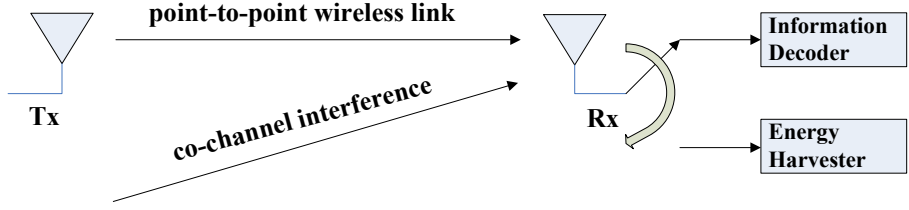}
\caption{System model.} \label{fig1}
\end{figure}

As shown in Fig. \ref{fig1}, this paper considers a wireless
point-to-point link consisting of one pair of single-antenna
transmitter (Tx) and receiver (Rx) over the flat-fading channel. It
is assumed that there is an aggregate interference at Rx, which is
within the same bandwidth as the transmitted signal from Tx, and
changes over time. For convenience, we assume that the channel from
Tx to Rx follows a block-fading model \cite{Shamai98}. Since the
coherence time for the time-varying interference is in general different from the
channel coherence time, we choose the block duration to be sufficiently small as compared to the minimum coherence time of the channel and interference such that they are both assumable to be constant during each block transmission. It is worth noting that the above
model is an example of the ``block interference'' channel introduced
in \cite{Stark}. The channel power gain and the interference power at Rx for
one particular fading state are denoted by $h(\nu)$ and $I(\nu)$,
respectively, where $\nu$ denotes the joint fading state. It is assumed that $h(\nu)$ and $I(\nu)$ are two random variables (RVs) with a joint probability density function (PDF) denoted by $f_\nu(h,I)$. At any fading state $\nu$,
$h(\nu)$ and $I(\nu)$ are assumed to be perfectly known at Rx. In
addition, the additive noise at Rx is assumed to be a circularly
symmetric complex Gaussian (CSCG) RV with zero mean and variance
$\sigma^2$.

We consider block-based transmissions at Tx and the time-switching
scheme \cite{Rui11} at Rx for decoding information or harvesting
energy at each fading state. Next, we elaborate the encoding and
decoding strategies for our system of interest in the following two
cases: Case I: $h(\nu)$ and $I(\nu)$ are unknown at Tx for all the
fading states of $\nu$, referred to as \emph{CSI Unknown at Tx}; and
Case II: $h(\nu)$ and $I(\nu)$ are perfectly known at Tx at each
fading state $\nu$, referred to as \emph{CSI Known at Tx} (CSIT).

\begin{figure}
\centering
 \epsfxsize=0.8\linewidth
    \includegraphics[width=12cm]{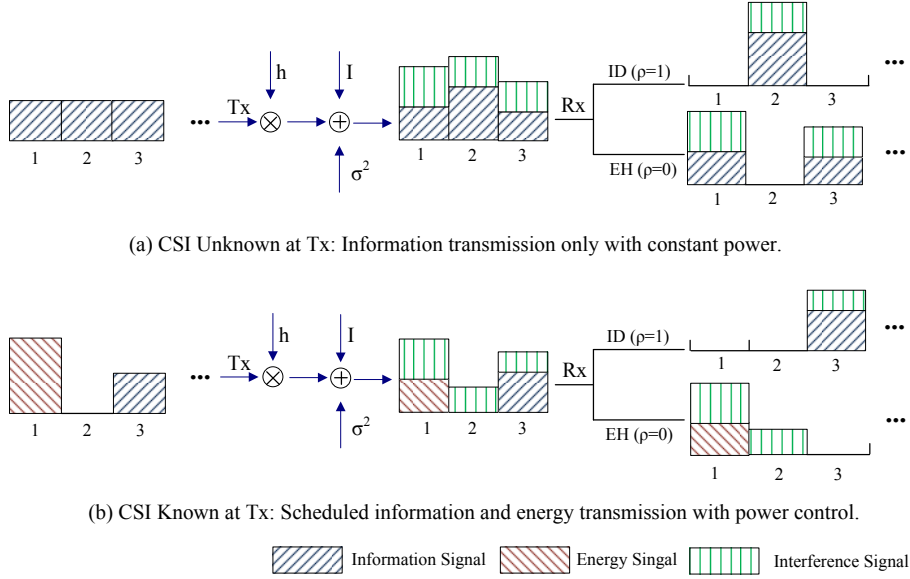}
\caption{Encoding and decoding strategies for wireless information
transfer with opportunistic WEH (via receiver mode switching). The
height of the block shown in the figure denotes the signal power.}
\label{fig2}
\end{figure}

First, consider the case of CSI Unknown at Tx. As shown in Fig.
\ref{fig2}(a), in this case Tx transmits information continuously
with constant power $P$ for all the fading states due to the lack of
CSIT. At each fading state $\nu$, Rx decides whether to decode the
information or harvest the energy from the received signal based on
$h(\nu)$ and $I(\nu)$. For example, as shown in Fig. \ref{fig2}(a),
time slots 1 and 3 are switched to EH mode at Rx, while time slot 2
is switched to ID mode. For convenience, we define an indicator
function to denote the receiver's mode switching at any given $\nu$
as follows:\begin{align}\label{eqn:indicator
function}\rho(\nu)=\left\{\begin{array}{ll}1, & {\rm ID \ mode \ is
\ active}
\\ 0, & {\rm EH \ mode \ is \ active}. \end{array} \right. \end{align}

Next, we consider the case of CSI Known at Tx, i.e., the channel
gain $h(\nu)$ and interference power $I(\nu)$ are known at Tx for
each fading state $\nu$. In this case, Tx is able to schedule
transmission for information and energy transfer to Rx based on the
instantaneous CSI. As shown in Fig. \ref{fig2}(b), Tx allocates time
slot 1 for energy transfer, time slot 3 for information transfer,
and transmits no signals in time slot 2. Accordingly, Rx will be in
EH mode (i.e., $\rho(\nu)=0$) to harvest energy from the received
signal (including the interference) in time slot 1 or solely from
the received interference in time slot 2, but in ID mode (i.e.,
$\rho(\nu)=1$) to decode the information in time slot 3. In addition
to transmission scheduling, Tx can implement power control based on
the CSI to further improve the information/energy transmission
efficiency. Let $p(\nu)$ denote the transmit power of Tx at fading
state $\nu$. In this paper, we consider two types of power
constraints on $p(\nu)$, namely {\it average power constraint} (APC)
and {\it peak power constraint} (PPC) \cite{Shamai98}. The APC
limits the average transmit power of Tx over all the fading states,
i.e., $E_{\nu}[p(\nu)]\leq P_{{\rm avg}}$, where $E_{\nu}[\cdot]$
denotes the expectation over $\nu$. In contrast, the PPC constrains
the instantaneous transmit power of Tx at each of the fading states,
i.e., $p(\nu)\leq P_{{\rm peak}}$, $\forall \nu$. Without loss of
generality, we assume $P_{{\rm avg}}\leq P_{{\rm peak}}$. For
convenience, we define the set of feasible power allocation as
\begin{align}\label{eqn:feasible power
set}\mathcal{P}\triangleq\big\{p(\nu): E_{\nu}[p(\nu)]\leq P_{{\rm
avg}}, p(\nu)\leq P_{{\rm peak}}, \forall \nu\big\}.\end{align}

\section{Information Transfer and Energy Harvesting Trade-offs in Fading Channels}\label{sec:Information Transfer verse Energy Harvesting Trade-offs in Fading Channels}

In this paper, we consider three performance measures at Rx, which
are the outage probability and the ergodic capacity for wireless
information transfer and the average harvested energy for WEH. For
delay-limited information transmission, outage probability is a
relevant performance indicator. Assuming that the interference is
treated as additive Gaussian noise at Rx and the transmitted signal
is Gaussian distributed, the instantaneous mutual information (IMI)
for the Tx-Rx link at fading state $\nu$ is expressed as
\begin{align}\label{eqn:rate new}r(\nu)=\rho(\nu)
\log\left(1+\frac{h(\nu)p(\nu)}{I(\nu)+\sigma^2}\right).\end{align}Note
that $r(\nu)=0$ if Rx switches to EH mode (i.e., $\rho(\nu)=0)$.
Thus, considering a delay-limited transmission with constant rate
$r_0$, following \cite{Shamai94} the outage probability at Rx can be
expressed as \begin{align}\label{eqn:new outage
probability}\varepsilon=Pr\left\{r(\nu)<
r_0\right\},\end{align}where $Pr\{\cdot\}$ denotes the probability.
For information transfer without CSIT, the receiver-aware outage
probability is usually minimized with a constant transmit power,
i.e., $p(\nu)=P_{{\rm avg}}\triangleq P$, $\forall \nu$
\cite{Shamai94}, whereas in the case with CSIT, the
transmitter-aware outage probability can be further minimized with
the ``truncated channel inversion'' based power allocation
\cite{Caire}, \cite{Goldsmith}.

Next, consider the case of no-delay-limited information transmission
for which the ergodic capacity is a suitable performance measure expressed as
\begin{align}\label{eqn:ergodic
capacity}R=E_\nu[r(\nu)].\end{align}For information transfer, if
CSIT is not available, the ergodic capacity can be achieved by a
random Gaussian codebook with constant transmit power over all
different fading states \cite{Shamai99}; however, with CSIT, the
ergodic capacity can be further maximized by the ``water-filling''
based power allocation \cite{Goldsmith}.

On the other hand, the amount of energy (normalized to the
transmission block duration) that can be harvested at Rx at fading
state $\nu$ is expressed as
$Q(\nu)=\alpha\big(1-\rho(\nu)\big)\big(h(\nu)p(\nu)+I(\nu)+\sigma^2\big)$,
where $\alpha$ is a constant that accounts for the loss in the
energy transducer for converting the harvested energy to electrical
energy to be stored; for convenience, it is assumed that $\alpha=1$
in this paper. Moreover, since the background thermal noise has
constant power $\sigma^2$ for all the fading states and $\sigma^2$
is typically a very small amount for energy harvesting, we may
ignore it in the expression of $Q(\nu)$. Thus, in the rest of this paper, we
assume
\begin{align}\label{eqn:energy new}Q(\nu):=\big(1-\rho(\nu)\big)\big(h(\nu)p(\nu)+I(\nu)\big).\end{align}
The average energy that can be harvested at Rx is then given
by\begin{align}\label{eqn:harvested energy}Q_{{\rm
avg}}=E_\nu[Q(\nu)].\end{align}

It is easy to see that there exist non-trivial trade-offs in
assigning the receiver mode $\rho(\nu)$ and/or transmit power
$p(\nu)$ (in the case of CSIT) to balance between minimizing the
outage probability or maximizing the ergodic capacity for
information transfer versus maximizing the average harvested energy
for WEH. To characterize such trade-offs, for the case when
information transmission is delay-limited, we introduce a so-called
{\it Outage-Energy} (O-E) region (defined below) that consists of
all the achievable non-outage probability (defined as
$\delta=1-\varepsilon$ with outage probability $\varepsilon$ given
in (\ref{eqn:new outage probability})) and average harvested energy
pairs for a given set of transmit power constraints, while for the
case when information transmission is not delay-limited, we use
another {\it Rate-Energy} (R-E) region (defined below) that consists
of all the achievable ergodic capacity and average harvested energy
pairs. More specifically, in the case without (w/o) CSIT, the
corresponding O-E region is defined
as\begin{align}\label{eqn:outage-energy region without
CSI}\mathcal{C}_{{\rm O-E}}^{{\rm w/o \ CSIT}}\triangleq
\bigcup\limits_{\rho(\nu)\in\{0,1\}, \forall \nu} \bigg\{
(\delta,Q_{{\rm avg}}):\delta \leq Pr\left\{r(\nu)\geq r_0\right\},
Q_{{\rm avg}}\leq E_\nu\left[Q(\nu)\right]\bigg\},\end{align}while
in the case with CSIT, the O-E region is defined
as\begin{align}\label{eqn:outage-energy region with
CSI}\mathcal{C}_{{\rm O-E}}^{{\rm with \ CSIT}}\triangleq
\bigcup\limits_{p(\nu)\in\mathcal{P}, \rho(\nu)\in\{0,1\}, \forall
\nu} \bigg\{(\delta,Q_{{\rm avg}}):\delta \leq Pr\left\{r(\nu)\geq
r_0\right\},  Q_{{\rm avg}}\leq
E_\nu\left[Q(\nu)\right]\bigg\}.\end{align}

On the other side, in the case without CSIT, the R-E region is defined
as\begin{align}\label{eqn:rate-energy region without
CSI}\mathcal{C}_{{\rm E-E}}^{{\rm w/o \ CSIT}}\triangleq
\bigcup\limits_{\rho(\nu)\in\{0,1\}, \forall \nu} \bigg\{ (R,Q_{{\rm
avg}}):R \leq E_\nu[r(\nu)], Q_{{\rm avg}}\leq
E_\nu\left[Q(\nu)\right]\bigg\},\end{align}while in the case with CSIT, the R-E region is defined
as\begin{align}\label{eqn:rate-energy region with
CSI}\mathcal{C}_{{\rm R-E}}^{{\rm with \ CSIT}}\triangleq
\bigcup\limits_{p(\nu)\in\mathcal{P}, \rho(\nu)\in\{0,1\}, \forall
\nu} \bigg\{(R,Q_{{\rm avg}}):R \leq E_\nu[r(\nu)], Q_{{\rm
avg}}\leq E_\nu\left[Q(\nu)\right]\bigg\}.\end{align}

\begin{figure}
\begin{center}
\subfigure[O-E
region]{\scalebox{0.4}{\includegraphics*{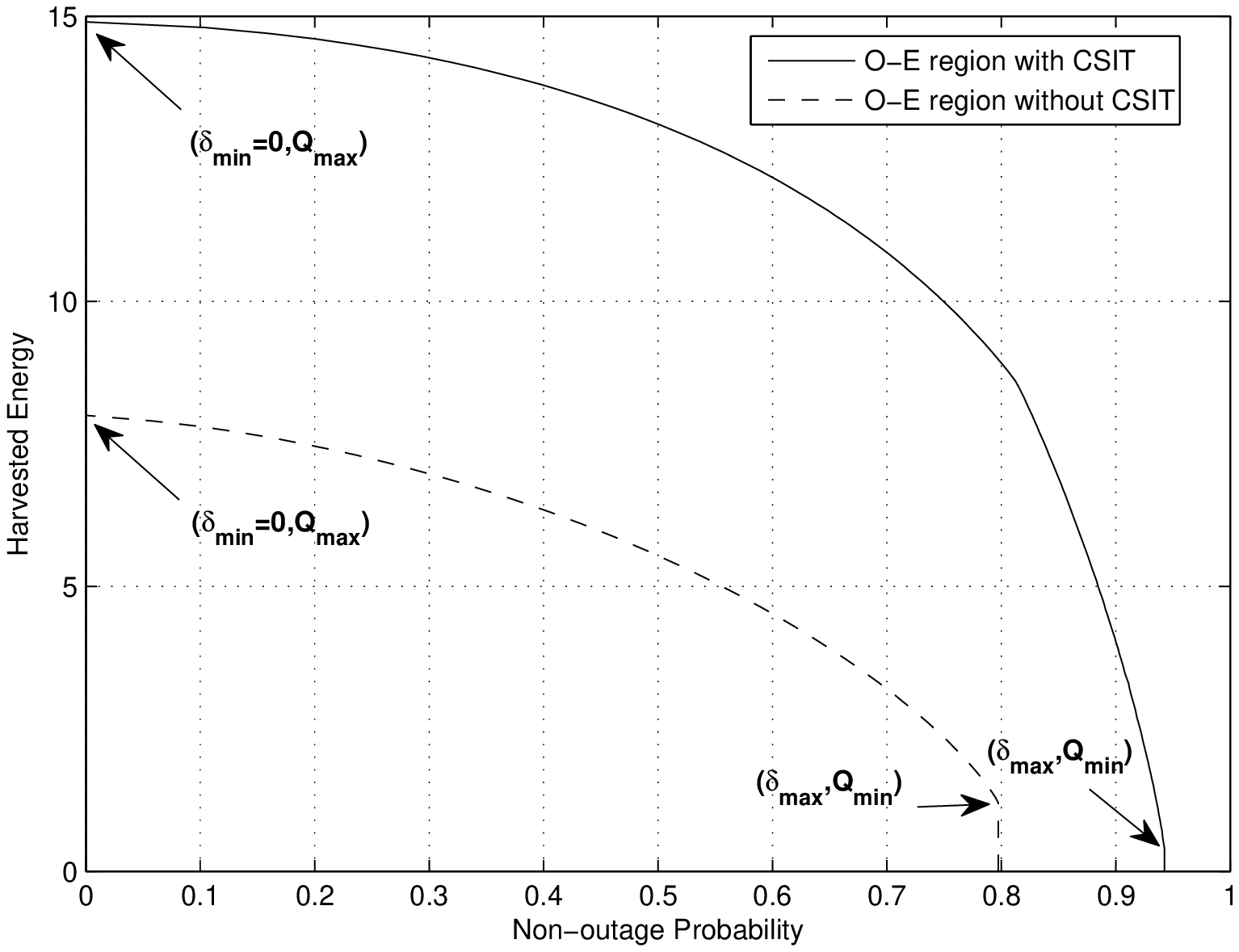}}}
\subfigure[R-E
region]{\scalebox{0.4}{\includegraphics*{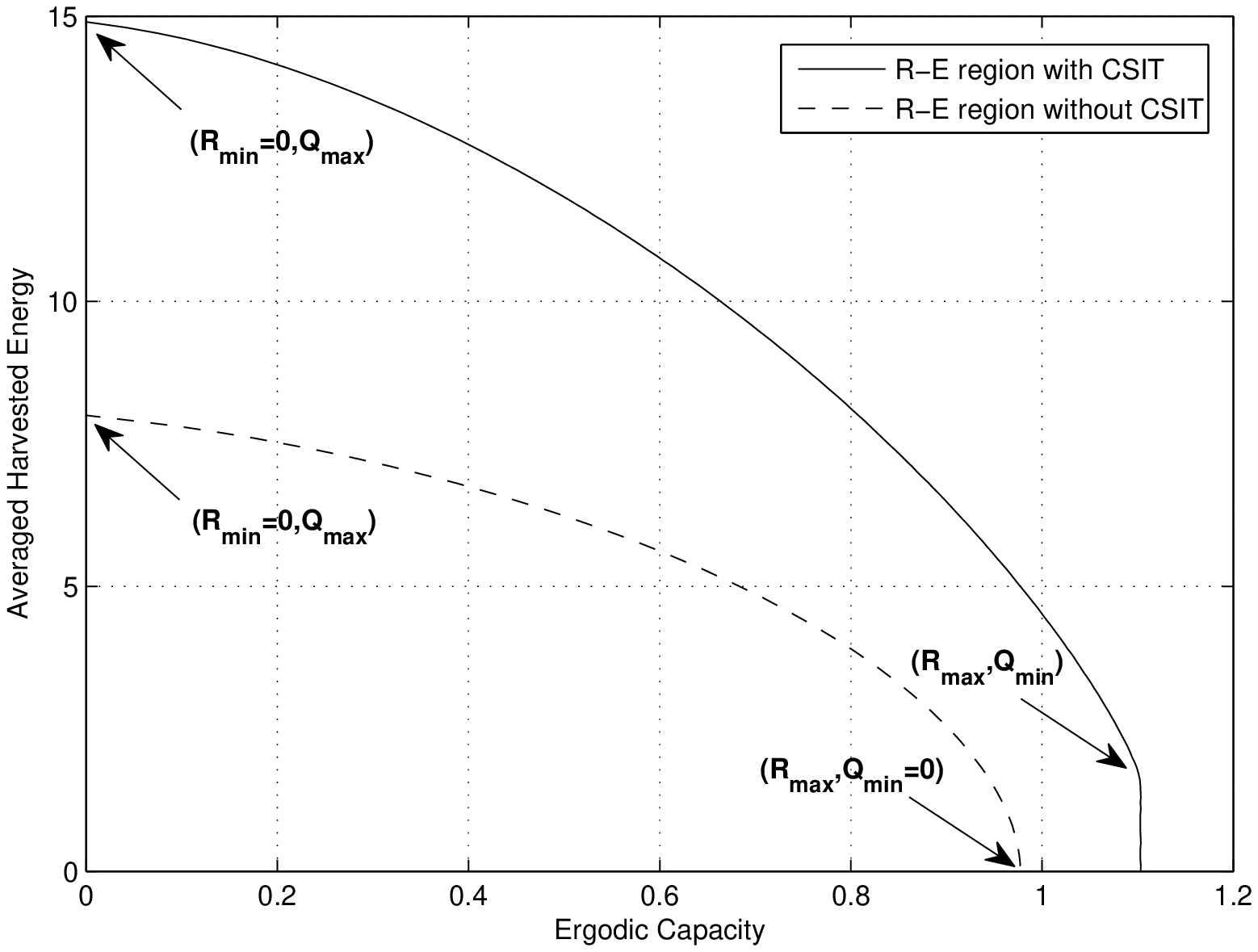}}}
\caption{Examples of O-E region and R-E region with or without
CSIT.}\label{fig3}
\end{center}

\end{figure}

Fig. \ref{fig3}(a) and Fig. \ref{fig3}(b) show examples of the O-E
region without or with CSIT (see Sections \ref{sec:Optimal Solution
to Problem (P1)} and \ref{sec:Optimal Solution to Problem (P2)} for
the details of computing the O-E regions for these two cases) and
the R-E region without or with CSIT (see Sections \ref{sec:Optimal
Solution to Problem (P3)} and \ref{sec:Optimal Solution to Problem
(P4)} for the corresponding details), respectively. It is assumed
that $P_{{\rm avg}}=5$, $P_{{\rm peak}}=20$, $\sigma^2=0.5$,
$r_0=0.3$, $h(\nu)$ and $I(\nu)$ are independent exponentially
distributed RVs with mean $1$ and $3$, respectively. It is observed
that CSIT helps improve both the achievable outage-energy and
rate-energy trade-offs.

It is observed from Fig. \ref{fig3} that in each region, there are
two boundary points that indicate the extreme performance limits,
namely, $(\delta_{{\rm max}},Q_{{\rm min}})$ and $(\delta_{{\rm
min}},Q_{{\rm max}})$ for the O-E region, or $(R_{{\rm max}},Q_{{\rm
min}})$ and $(R_{{\rm min}},Q_{{\rm max}})$ for the R-E region. For
brevity, characterizations of these vertex points are given in
Appendix.

%
%

Since the optimal trade-offs between the non-outage
probability/ergodic capacity and the average harvested energy are
characterized by the boundary of the corresponding O-E/R-E region,
it is important to characterize all the boundary $(\delta,Q_{{\rm
avg}})$ or $(R,Q_{{\rm avg}})$ pairs in each case with or without
CSIT. From Fig. \ref{fig3}, it is easy to observe that if $Q_{{\rm
avg}}<Q_{{\rm min}}$, the non-outage probability $\delta_{{\rm
max}}$ or ergodic capacity $R_{{\rm max}}$ can still be achieved for
both cases with and without CSIT. Thus, the remaining boundary of
the O-E region yet to be characterized is over the intervals
$Q_{{\rm min}}\leq Q_{{\rm avg}}\leq Q_{{\rm max}}$ and
$\delta_{{\rm min}}\leq \delta \leq \delta_{{\rm max}}$, while that
of the R-E region is over the intervals $Q_{{\rm min}}\leq Q_{{\rm
avg}}\leq Q_{{\rm max}}$ and $R_{{\rm min}}\leq R \leq R_{{\rm
max}}$.

For the O-E region, we introduce the following indicator function
for the event of non-outage transmission at fading state $\nu$ for
the convenience of our subsequent analysis:\begin{align}\label{eqn:x}
X(\nu)=\left\{\begin{array}{ll}1, & {\rm if} \ r(\nu)\geq r_0 \\ 0,
& {\rm otherwise}.
\end{array}\right.\end{align}It thus follows that the non-outage
probability $\delta$ can be reformulated
as\begin{align}\label{eqn:new out}\delta=Pr\{r(\nu)\geq
r_0\}=E_\nu[X(\nu)].\end{align}Then, we consider the following two
optimization
problems.\begin{align*}\mathrm{(P1)}:~\mathop{\mathtt{Maximize}}_{\{\rho(\nu)\}}
& ~~~ E_\nu[X(\nu)] \\
\mathtt {Subject \ to} & ~~~ E_\nu[Q(\nu)]\geq \bar{Q} \\ & ~~~
\rho(\nu)\in \{0,1\}, \ \forall \nu
\end{align*}
\begin{align*}\mathrm{(P2)}:~\mathop{\mathtt{Maximize}}_{\{p(\nu),\rho(\nu)\}}
& ~~~ E_\nu[X(\nu)] \\
\mathtt {Subject \ to} & ~~~ E_\nu[Q(\nu)]\geq \bar{Q} \\
& ~~~ p(\nu) \in \mathcal{P}, \ \forall \nu \\ & ~~~ \rho(\nu)\in
\{0,1\}, \ \forall \nu
\end{align*}where $\bar{Q}$ is a target average harvested energy required to maintain the receiver's operation. By solving
Problem (P1) or (P2) for all $Q_{{\rm min}}\leq \bar{Q}\leq Q_{{\rm
max}}$, we are able to characterize the entire boundary of the O-E
region for the case without CSIT (defined in (\ref{eqn:outage-energy
region without CSI})) or with CSIT (defined in
(\ref{eqn:outage-energy region with CSI})).

Similarly, for the R-E region, we consider the following two optimization
problems.\begin{align*}\mathrm{(P3)}:~\mathop{\mathtt{Maximize}}_{\{\rho(\nu)\}}
& ~~~ E_\nu[r(\nu)] \\
\mathtt {Subject \ to} & ~~~ E_\nu[Q(\nu)]\geq \bar{Q} \\ & ~~~
\rho(\nu)\in \{0,1\}, \ \forall \nu
\end{align*}
\begin{align*}\mathrm{(P4)}:~\mathop{\mathtt{Maximize}}_{\{p(\nu),\rho(\nu)\}}
& ~~~ E_\nu[r(\nu)] \\
\mathtt {Subject \ to} & ~~~ E_\nu[Q(\nu)] \geq \bar{Q} \\ & ~~~
p(\nu) \in \mathcal{P}, \ \forall \nu \\ & ~~~ \rho(\nu)\in \{0,1\},
\ \forall \nu.
\end{align*}Then, by solving Problem (P3) or (P4) for all $Q_{{\rm min}}\leq \bar{Q}\leq Q_{{\rm
max}}$, we can characterize the boundary of the R-E region for the
case without CSIT (defined in (\ref{eqn:rate-energy region without
CSI})) or with CSIT (defined in (\ref{eqn:rate-energy region with
CSI})).

It is observed that the objective function of Problem (P2)
is in general not concave in $p(\nu)$ even if $\rho(\nu)$'s are given. Furthermore, due to the
integer constraint $\rho(\nu)\in\{0,1\}$, $\forall \nu$, Problems
(P1)-(P4) are in general non-convex optimization problems. However,
it can be verified that all of them satisfy the
``time-sharing'' condition given in \cite{Yu06}. To show this for
Problem (P1), let $\Phi_1(\bar{Q})$ denote the optimal problem value
given the harvested energy constraint $\bar{Q}$, and
$\{\rho^a(\nu)\}$ and $\{\rho^b(\nu)\}$ denote the optimal solutions
given the harvested energy constraints $\bar{Q}^a$ and $\bar{Q}^b$,
respectively. We need to prove that for any $0\leq \theta \leq 1$,
there always exists at least one solution $\{\rho^c(\nu)\}$ such
that $E_\nu[X^c(\nu)]\geq \theta
\Phi_1(\bar{Q}^a)+(1-\theta)\Phi_2(\bar{Q}^b)$ and $E_\nu[Q^c(\nu)]\geq
\theta \bar{Q}^a+(1-\theta)\bar{Q}^b$, where
$Q^c(\nu)=\big(1-\rho^c(\nu)\big)\big(h(\nu)P+I(\nu)\big)$ and
$X^c(\nu)$ is defined accordingly as in (\ref{eqn:x}). Due to the
space limitation, the above proof is omitted here. In fact, the
``time-sharing'' condition implies that $\Phi_1(\bar{Q})$ is concave
in $\bar{Q}$, which then guarantees the zero duality gap for Problem
(P1) according to the convex analysis in \cite{Rockafellar70}. Similarly, it can be shown that strong duality holds for Problems (P2)-(P4). Therefore, in the
following two sections, we apply the Lagrange duality method to
solve Problems (P1)-(P4) to obtain the optimal O-E and R-E
trade-offs, respectively.

\section{Outage-Energy Trade-off}\label{sec:Outage Probability verse Averaged Harvested
Energy}

In this section, we study the optimal receiver mode switching without/with
transmit power control to achieve different trade-offs between the
minimum outage probability and the maximum average harvested energy
for both cases without and with CSIT by solving Problems (P1) and
(P2), respectively.

\subsection{The Case Without CSIT: Optimal Receiver Mode Switching}\label{sec:Optimal Solution to Problem (P1)}

We first study Problem (P1) for the CSIT-unknown case to derive the
optimal rule at Rx to switch between EH and ID modes. The Lagrangian
of Problem (P1) is formulated as\begin{align}\label{eqn:Lagrangian1}
L(\rho(\nu),\lambda)=E_\nu[X(\nu)]+\lambda\left(E_\nu[Q(\nu)]-\bar{Q}\right),
\end{align}where $\lambda\geq 0$ is the dual
variable associated with the harvested energy constraint $\bar{Q}$.
Then, the Lagrange dual function of Problem (P1) is expressed as
\begin{align}\label{eqn:dual function}
g(\lambda)=\max\limits_{\rho(\nu)\in \{0,1\},\forall
\nu}L(\rho(\nu),\lambda).
\end{align}The maximization problem (\ref{eqn:dual
function}) can be decoupled into parallel subproblems all having the
same structure and each for one fading state. For a particular
fading state $\nu$, the associated subproblem is expressed as
\begin{align}\label{eqn:subproblem}\max_{\rho \in \{0,1\}}
~~~ L_{\nu}^{{\rm O-E}}(\rho),\end{align}where $L_{\nu}^{{\rm
O-E}}(\rho)=X+\lambda Q$. Note that we have dropped the index $\nu$
for the fading state for brevity.

To solve Problem (\ref{eqn:subproblem}), we need to compare the
values of $L_{\nu}^{{\rm O-E}}(\rho)$ for $\rho=1$ and $\rho=0$. It
follows from (\ref{eqn:energy new}), (\ref{eqn:x}) and
(\ref{eqn:Lagrangian1}) that when $\rho=1$,
\begin{align}\label{eqn:Lagrangian ID} L_{\nu}^{{\rm O-E}}(\rho=1)=\left\{\begin{array}{ll}1,
&{\rm if} \ \frac{h}{I+\sigma^2}>\frac{e^{r_0}-1}{P} \\ 0, & {\rm
otherwise}
\end{array}\right.\end{align}and when
$\rho=0$,\begin{align}\label{eqn:Lagrangian EH} L_{\nu}^{{\rm
O-E}}(\rho=0)=\lambda hP+\lambda I.
\end{align}Thus, the optimal
solution to Problem (\ref{eqn:subproblem}) is obtained
as\begin{align}\label{eqn:optimal
solution}\rho^\ast=\left\{\begin{array}{ll}1, & {\rm if} \
\frac{h}{I+\sigma^2}>\frac{e^{r_0}-1}{P} \ {\rm and} \ \lambda
hP+\lambda I<1 \\ 0, & {\rm otherwise}. \end{array} \right.
\end{align}

With a given $\lambda$, Problem (\ref{eqn:dual function}) can be
efficiently solved by solving Problem (\ref{eqn:subproblem}) for
different fading states. Problem (P1) is then solved by iteratively
solving Problem (\ref{eqn:dual function}) with a fixed $\lambda$,
and updating $\lambda$ via a simple bisection method until the
harvested energy constraint is met with equality \cite{Boyd04}.

Next, we examine the optimal solution $\rho^\ast$ to Problem (P1) to
gain more insights to the optimal receiver mode switching in the
case without CSIT. With a given harvested energy constraint
$\bar{Q}$, we define the region on the $(h,I)$ plane consisting of
all the points $(h,I)$ for which the optimal solution to Problem
(P1) is $\rho^\ast=1$ (versus $\rho^\ast=0$) as the optimal ID
region (versus the optimal EH region). Furthermore, let
$\lambda^\ast$ denote the optimal dual solution to Problem (P1)
corresponding to the given $\bar{Q}$. Then, from (\ref{eqn:optimal
solution}) the optimal ID region for Problem (P1) is expressed as
\begin{align}\label{eqn:optimal ID region}\mathcal{D}_{{\rm
ID}}(\lambda^\ast)\triangleq\bigg\{\big(h,I\big):
\frac{h}{I+\sigma^2}>\frac{e^{r_0}-1}{P},1>\lambda^\ast
hP+\lambda^\ast I, h>0,I>0 \bigg\}.\end{align}The rest of the
non-negative $(h,I)$ plane is thus the optimal EH region,
i.e.,\begin{align}\label{eqn:optimal EH region}\mathcal{D}_{{\rm
EH}}(\lambda^\ast)\triangleq \mathbb{R}^2_+\backslash
\mathcal{D}_{{\rm ID}}(\lambda^\ast),\end{align}where
$\mathbb{R}^2_+$ denotes the two-dimensional nonnegative real
domain, and $A \backslash B$ denotes the set $\{x|x\in A \ {\rm and}
\ x \not \in B\}$.

\begin{figure}
\centering
 \epsfxsize=0.7\linewidth
    \includegraphics[width=10cm]{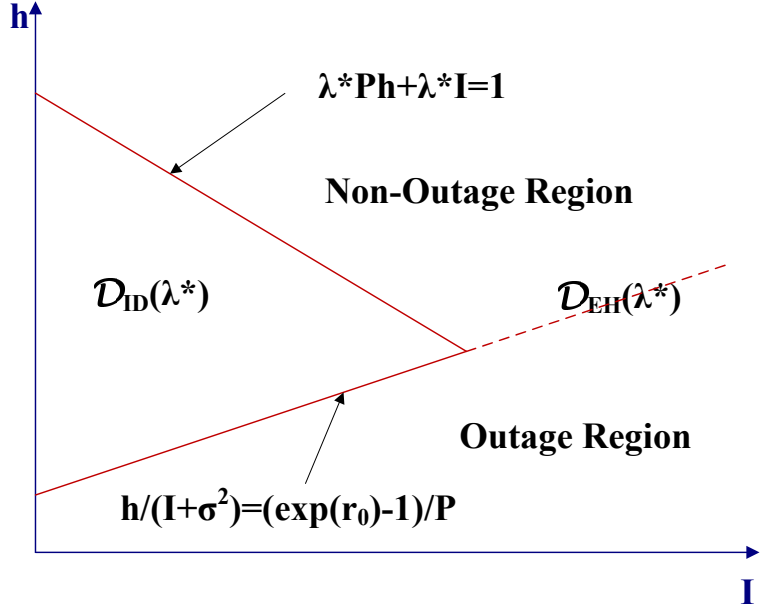}
\caption{Illustration of the optimal ID and EH regions for
characterizing O-E trade-offs in the case without CSIT.}
\label{fig4}
\end{figure}

An illustration of $\mathcal{D}_{{\rm ID}}(\lambda^\ast)$ and
$\mathcal{D}_{{\rm EH}}(\lambda^\ast)$ is shown in Fig. \ref{fig4}
with $\bar{Q}>Q_{{\min}}$. It is noted that to meet the harvested
energy constraint $\bar{Q}$, we need to sacrifice (increase) the
outage probability for information transfer by allocating some
non-outage fading states in the region
$\mathcal{H}=\{(h,I):\log\big(1+\frac{hP}{I+\sigma^2}\big)\geq
r_0\}$ to EH mode. An interesting question here is to decide which
portion of $\mathcal{H}$ should be allocated to EH mode. It is
observed from Fig. \ref{fig4} that the optimal way is to allocate
all $(h,I)$ pairs satisfying $1<\lambda^\ast hP+\lambda^\ast I$ or
$hP+I>\frac{1}{\lambda^\ast}$ in $\mathcal{H}$ to EH mode, i.e., the
fading states with sufficiently large signal plus interference total
power values at Rx should be allocated to EH mode. This is
reasonable since if we have to allocate a certain number of fading
states in $\mathcal{H}$ to EH mode, i.e., increase the transmission
outage probability by the same amount, these fading states should be
chosen to maximize the harvested energy at Rx.

Furthermore, note that $\lambda^\ast$ increases monotonically with
$\bar{Q}$. Thus, the boundary line $\lambda^\ast hP+\lambda^\ast
I=1$ that separates the optimal ID and EH regions in Fig. \ref{fig4}
will be shifted down as $\lambda^\ast$ increases, and as a result
$\mathcal{D}_{{\rm ID}}(\lambda^\ast)$ shrinks. It can be shown that
if $\lambda^\ast\geq \frac{1}{(e^{r_0}-1)\sigma^2}$, then
$\mathcal{D}_{{\rm ID}}(\lambda^\ast)=\O$, which corresponds to the
point $(\delta_{{\rm min}}=0,Q_{{\rm max}})$ of the O-E region shown in Fig. \ref{fig3}(a)
for the case without CSIT.

It is worth noting that if $I(\nu)=0$, $\forall \nu$, then the optimal ID region reduces to $\mathcal{D}_{{\rm
ID}}(\lambda^\ast)=\{h:\frac{(e^{r_0}-1)\sigma^2}{P}\leq h \leq \frac{1}{\lambda^\ast P}\}$, and the rest of the $h$-axis is thus the EH region. In this case, the outage fading states $h\in (0,\frac{(e^{r_0}-1)\sigma^2}{P})$ are all allocated to EH mode since they cannot be used by ID mode. However, the harvested energy in the outage states only accounts for a small portion of the total harvested energy due to the poor channel gains. Most of the energy is harvested in the interval $h\in (\frac{1}{\lambda^\ast P},\infty)$, i.e., when the channel power is above a certain threshold.

\subsection{The Case With CSIT: Joint Information/Energy Scheduling, Power Control, and
Receiver Mode Switching}\label{sec:Optimal Solution to Problem (P2)}
In this subsection, we address the case of CSI known at Tx and
jointly optimize the energy/information scheduling and power
control at Tx, as well as EH/ID mode switching at Rx, as formulated
in Problem (P2). Let $\lambda$ and $\beta$ denote the nonnegative
dual variables corresponding to the average harvested energy
constraint and average transmit power constraint, respectively.
Similarly as for Problem (P1), Problem (P2) can be decoupled into
parallel subproblems each for one particular fading state and
expressed as (by ignoring the fading index
$\nu$)\begin{align}\label{eqn:subproblem2}\max_{0\leq p\leq P_{{\rm
peak}},\rho \in \{0,1\}}  ~~~ L_{\nu}^{{\rm
O-E}}(p,\rho),\end{align}where $L_{\nu}^{{\rm
O-E}}(p,\rho)=X+\lambda Q-\beta p$. To solve Problem
(\ref{eqn:subproblem2}), we need to compare the optimal values of
$L_\nu^{{\rm O-E}}(p,\rho)$ for $\rho=1$ and $\rho=0$, respectively,
as shown next.

When $\rho=1$, it follows that
\begin{align}\label{eqn:id}L_{\nu}^{{\rm O-E}}(p,\rho=1)=\left\{\begin{array}{ll}1-\beta p,
& {\rm if} \ p\geq \bar{p} \\ -\beta p, & {\rm otherwise}
\end{array} \right. \end{align}where
$\bar{p}=\frac{(e^{r_0}-1)(I+\sigma^2)}{h}$. It can be verified that
the optimal power allocation for the ID mode to maximize
(\ref{eqn:id}) subject to $0\leq p \leq P_{{\rm peak}}$ is the
well-known ``truncated channel inversion'' policy \cite{Goldsmith}
given by\begin{align}\label{eqn:new channel inversion}p_{{\rm
ID}}=\left\{\begin{array}{ll}\bar{p}, & {\rm if} \ \frac{h}{I+\sigma^2}\geq h_1 \\
0, & {\rm otherwise} \end{array} \right. \end{align}where
$h_1=\max\{\beta(e^{r_0}-1),\frac{e^{r_0}-1}{P_{{\rm peak}}}\}$.

When $\rho=0$, it follows that
\begin{align}\label{eqn:rho=0}L_{\nu}^{{\rm O-E}}(p,\rho=0)=\lambda hp+\lambda
I-\beta p. \end{align}

Define $h_2=\frac{\beta}{\lambda}$. Then the optimal power
allocation for the EH mode can be expressed as
\begin{align}\label{eqn:optimal power for EH mode}p_{{\rm EH}}=\left\{\begin{array}{ll}P_{{\rm peak}},
& {\rm if} \ h\geq h_2 \\ 0, & {\rm otherwise}. \end{array} \right.
\end{align}

To summarize, we have\begin{align} & L_{\nu}^{{\rm O-E}}(p_{{\rm
ID}},\rho=1)=\left\{\begin{array}{ll}1-\beta \bar{p}, & {\rm if} \
\frac{h}{I+\sigma^2}\geq h_1 \\ 0, &
{\rm otherwise}; \end{array} \right.  \label{eqn:Lagrangian ID2} \\
 & L_{\nu}^{{\rm O-E}}(p_{{\rm EH}},\rho=0)=\left\{\begin{array}{ll}(\lambda
h-\beta) P_{{\rm peak}}+\lambda I, & {\rm if} \ h\geq h_2 \\
\lambda I, & {\rm otherwise}.
\end{array} \right.  \label{eqn:EH utility}\end{align}

Then, given any pair of $\lambda$ and $\beta$, the optimal solution
to Problem (\ref{eqn:subproblem2}) for fading state $\nu$ can be
expressed as\begin{eqnarray}&& \rho^\ast=\left\{\begin{array}{ll}1,
& {\rm if}
\ L_\nu^{{\rm O-E}}(p_{{\rm ID}},\rho=1)> L_\nu^{{\rm O-E}}(p_{{\rm EH}},\rho=0) \\
0, & {\rm otherwise}; \end{array}\right. \label{eqn:optimal solution 2}\\
&& p^\ast=\left\{\begin{array}{ll}p_{{\rm ID}}, & {\rm if} \
\rho^\ast=1 \\ p_{{\rm EH}}, & {\rm if} \ \rho^\ast=0.
\end{array} \right. \end{eqnarray}

Next, to find the optimal dual variables $\lambda^\ast$ and
$\beta^\ast$ for Problem (P2), sub-gradient based methods such as
the ellipsoid method \cite{Boyd04} can be applied. It can be shown
that the sub-gradient for updating $(\lambda,\beta)$ is
$[E_\nu[Q^\ast(\nu)]-\bar{Q},P_{{\rm avg}}-E_\nu[p^\ast(\nu)]]$,
where $Q^\ast(\nu)$ and $p^\ast(\nu)$ denote the harvested energy
and transmit power at fading state $\nu$, respectively, after
solving Problem (\ref{eqn:subproblem2}) for a given pair of
$\lambda$ and $\beta$. Hence, Problem (P2) is solved.

Next, we investigate further the optimal information/energy transfer scheduling and power control at Tx, as well as the optimal mode switching at Rx. For simplicity, we only study the case of $I(\nu)=0$, $\forall \nu$. From the above analysis, it follows that there are three possible
transmission modes at Tx for the case with CSIT: ``information transfer mode'' with channel inversion
power control, ``energy transfer mode'' with peak transmit power,
and ``silent mode'' with no transmission, where the first transmission
mode corresponds to ID mode at Rx and the second transmission
mode corresponds to EH mode at Rx. We thus define $\mathcal{B}_{{\rm on}}^{{\rm
ID}}$, $\mathcal{B}_{{\rm on}}^{{\rm
EH}}$, and $\mathcal{B}_{{\rm off}}$ on the non-negative $h$-axis as
the regions corresponding to the above three modes, respectively. Since the explicit expressions for characterizing these regions are complicated and depend on the values of $\bar{Q}$ and $P_{{\rm avg}}$, in the following we will study $\mathcal{B}_{{\rm on}}^{{\rm
ID}}$, $\mathcal{B}_{{\rm on}}^{{\rm
EH}}$, and $\mathcal{B}_{{\rm off}}$ in the special case of $h_1\geq h_2$ to shed some light on the optimal design. Let $\lambda^\ast$ and $\beta^\ast$ denote the optimal dual solutions to Problem (P2). With $h_1\geq h_2$, it can be shown that $\mathcal{B}_{{\rm on}}^{{\rm
ID}}=\{h:h_1\leq h \leq h_3\}$, $\mathcal{B}_{{\rm on}}^{{\rm
EH}}=\{h:h>h_3\}$ and $\mathcal{B}_{{\rm off}}=\{h:h<h_1\}$, where $h_3$ is the largest root of the equation: $\lambda^\ast P_{{\rm peak}}h^2-(\beta^\ast P_{{\rm peak}}+1)h+\beta^\ast(e^{r_0}-1)\sigma^2=0$. The proof is omitted here due to the space limitation.

\begin{figure}
\centering
 \epsfxsize=0.8\linewidth
    \includegraphics[width=10cm]{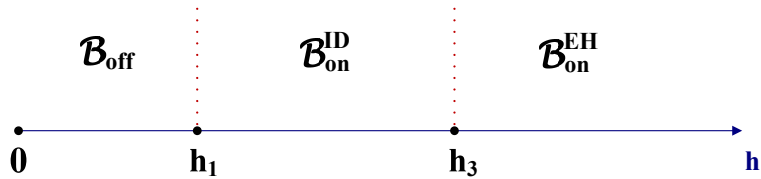}
\caption{Illustration of the optimal transmitter and receiver modes
for characterizing O-E trade-offs in the case with
CSIT. It is assumed that $I(\nu)=0$, $\forall \nu$, and $h_1\geq h_2$.}\label{fig5}
\end{figure}

An illustration of $\mathcal{B}_{{\rm on}}^{{\rm
ID}}$, $\mathcal{B}_{{\rm on}}^{{\rm
EH}}$, and $\mathcal{B}_{{\rm off}}$ for the case of $I(\nu)=0$, $\forall \nu$, and $h_1\geq h_2$ is shown in Fig.\ref{fig5}. Similar to the case without CSIT (cf. Fig. \ref{fig4}), the optimal design for the case with CSIT is still to allocate the best channels to the EH mode rather than the ID mode. However, unlike the case without CSIT, when the channel condition is poor, the transmitter in the case with CSIT will shut down its transmission to save power.

\section{Rate-Energy Trade-off}\label{sec:Ergodic Capacity verse Averaged Harvested Energy}

In this section, we investigate the optimal resource allocation
schemes to achieve different trade-offs between the maximum ergodic
capacity and maximum averaged harvested energy for the two cases
without and with CSIT by solving Problems (P3) and (P4),
respectively.

\subsection{The Case Without
CSIT: Optimal Receiver Mode Switching}\label{sec:Optimal Solution to
Problem (P3)}

First, we study Problem (P3) for the CSIT-unknown case to derive the
optimal switching rule at Rx between EH and ID modes for
characterizing different R-E trade-offs. Similarly as in Section
\ref{sec:Optimal Solution to Problem (P1)}, Problem (P3) can be
decoupled into parallel subproblems each for one particular fading
state $\nu$, expressed as
\begin{align}\label{eqn:subproblem3}\max_{\rho \in \{0,1\}}
~~~ L_{\nu}^{{\rm R-E}}(\rho),\end{align}where $L_{\nu}^{{\rm
R-E}}(\rho)=r+\lambda Q$ with $\lambda\geq 0$ denoting the dual
variable associated with the harvested energy constraint $\bar{Q}$.
Note that we have dropped the index $\nu$ of the fading state for
brevity.

To solve Problem (\ref{eqn:subproblem3}), we need to compare the
values of $L_{\nu}^{{\rm R-E}}(\rho)$ for $\rho=1$ and $\rho=0$.
When $\rho=1$, it follows that
\begin{align}\label{eqn:Lagrangian ID3} L_{\nu}^{{\rm
R-E}}(\rho=1)=\log\left(1+\frac{hP}{I+\sigma^2}\right).\end{align}When
$\rho=0$, it follows that\begin{align}\label{eqn:Lagrangian EH3}
L_{\nu}^{{\rm R-E}}(\rho=0)=\lambda hP+\lambda I.
\end{align}Thus, the optimal
solution to Problem (\ref{eqn:subproblem3}) is obtained
as\begin{align}\label{eqn:optimal
solution}\rho^\ast=\left\{\begin{array}{ll}1, & {\rm if} \ \log\big(1+\frac{hP}{I+\sigma^2}\big)>\lambda hP+\lambda I \\
0, & {\rm otherwise}. \end{array} \right.
\end{align}

To find the optimal dual variable $\lambda^\ast$ to Problem (P3), a
simple bisection method can be applied until the harvested energy
constraint is met with equality. Thus, Problem (P3) is efficiently
solved.

Similar to Section \ref{sec:Optimal Solution to Problem (P1)}, in
the following we characterize the optimal ID region and EH region to
get more insights to the optimal receiver mode switching for
characterizing different R-E trade-offs. Let $\lambda^\ast$ denote the optimal
dual variable corresponding to a given energy target $\bar{Q}$. The
optimal ID region can then be expressed
as\begin{align}\label{eqn:optimal ID region E-E}\mathcal{D}_{{\rm
ID}}(\lambda^\ast)\triangleq\bigg\{\big(h,I\big):
\log\left(1+\frac{hP}{I+\sigma^2}\right)>\lambda^\ast hP+\lambda^\ast I
\bigg\}.\end{align}The rest of the non-negative $(h,I)$ plane is
thus the optimal EH region, i.e.,\begin{align}\label{eqn:optimal EH
region E-E}\mathcal{D}_{{\rm EH}}(\lambda^\ast)\triangleq
\mathbb{R}^2_+\backslash \mathcal{D}_{{\rm
ID}}(\lambda^\ast).\end{align}

\begin{figure}
\centering
 \epsfxsize=0.6\linewidth
    \includegraphics[width=10cm]{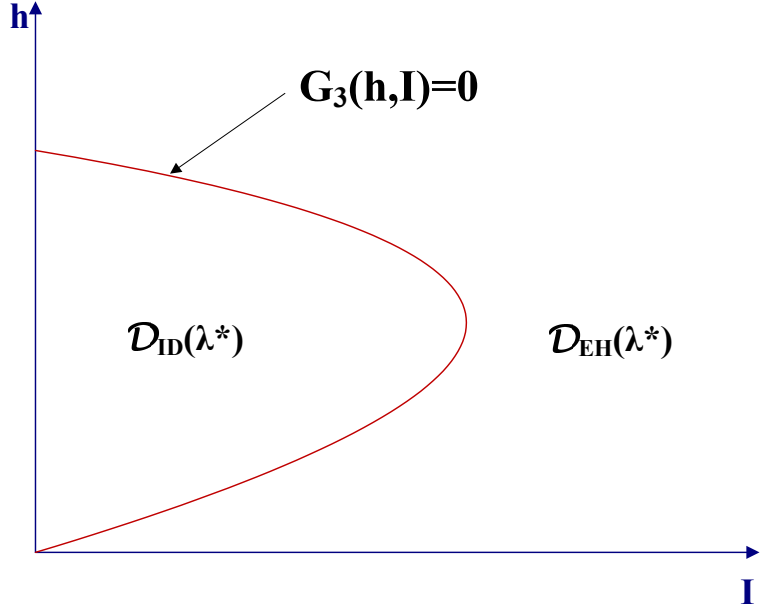}
\caption{Illustration of the optimal ID and EH regions for
characterizing R-E trade-offs in the case without CSIT.}
\label{fig6}
\end{figure}

Define $G_3(h,I)=\log\big(1+\frac{hP}{I+\sigma^2}\big)-(\lambda^\ast
hP+\lambda^\ast I)$. Fig. \ref{fig6} gives an illustration of the
optimal ID region and EH region for a particular value of
$\bar{Q}>Q_{{\rm min}}$.

Next, we discuss the optimal mode switching rule at Rx for achieving
various R-E trade-offs in the case without CSIT. Similar to the case
of O-E trade-off, for meeting the harvested energy constraint
$\bar{Q}$, we need to sacrifice (decrease) the ergodic capacity for
information transfer by allocating some fading states to EH mode.
Similar to the discussions in Section \ref{sec:Outage Probability verse
Averaged Harvested Energy}, the optimal rule is to allocate fading
states with largest values of $h$ for information transfer to EH
mode. The reason is that although fading states with good direct
channel gains are most desirable for ID mode, from
(\ref{eqn:Lagrangian ID3}) and (\ref{eqn:Lagrangian EH3}) it is
observed that the Lagrangian value of ID mode increases
logarithmically with $h$, while that of EH mode increases linearly
with $h$. As a result, when $h$ is above a certain threshold, the
value of $L_{\nu}^{{\rm R-E}}(\rho=0)$ will be larger than that of
$L_{\nu}^{{\rm R-E}}(\rho=1)$. In other words, when $h$ is good
enough, we can gain more by switching from ID mode to EH mode.

It is also observed that as the value of $\lambda^\ast$ increases,
the optimal ID region shrinks. In the following, we derive the value
of $\lambda^\ast$ corresponding to the point $(R_{{\rm
min}}=0,Q_{{\rm max}})$ in Fig. \ref{fig3}(b). From Fig. \ref{fig6}
it can be observed that $G_3(h,I)$ has two intersection points with
the $h$-axis, one of which is $(0,0)$. It can be shown that
$G_3(h,I=0)=\log\big(1+\frac{hP}{\sigma^2}\big)-\lambda^\ast hP$ is
a monotonically increasing function of $h$ in the interval
$(0,\frac{\frac{1}{\lambda^\ast}-\sigma^2}{P}]$, and decreasing
function of $h$ in the interval
$(\frac{\frac{1}{\lambda^\ast}-\sigma^2}{P},\infty)$. Consequently,
if $\frac{\frac{1}{\lambda^\ast}-\sigma^2}{P}=0$, i.e.,
$\lambda^\ast=\frac{1}{\sigma^2}$, the other intersection point of
$G_3(h,I)$ with the $h$-axis will coincide with the point $(0,0)$, and
thus $\mathcal{D}_{{\rm ID}}(\lambda^\ast)=\O$ if $\lambda^\ast\geq
\frac{1}{\sigma^2}$.

\subsection{The Case With CSIT: Joint Information/Energy Scheduling, Power Control, and
Receiver Mode Switching}\label{sec:Optimal Solution to Problem (P4)}
In this subsection, we study Problem (P4) to achieve different
optimal R-E trade-offs for the case of CSIT by jointly optimizing
energy/information scheduling and power control at Tx, together with
the EH/ID mode switching at Rx. For Problem (P4), let $\lambda$ and
$\beta$ denote the nonnegative dual variables corresponding to the
average harvested energy constraint and average transmit power
constraint, respectively. Then, Problem (P4) can be decoupled into
parallel subproblems each for one particular fading state and
expressed as (by ignoring the fading index
$\nu$)\begin{align}\label{eqn:subproblem4}\max_{0\leq p\leq P_{{\rm
peak}},\rho \in \{0,1\}}  ~~~ L_{\nu}^{{\rm
R-E}}(p,\rho),\end{align}where $L_{\nu}^{{\rm
R-E}}(p,\rho)=r+\lambda Q-\beta p$. To solve Problem
(\ref{eqn:subproblem4}), we need to compare the maximum values of
$L_\nu^{{\rm R-E}}(p,\rho)$ for $\rho=1$ and $\rho=0$, respectively,
as shown next.

When $\rho=1$, it follows that
\begin{align}L_\nu^{{\rm
R-E}}(p,\rho=1)=\log\left(1+\frac{hp}{I+\sigma^2}\right)-\beta
p.\end{align}It can be shown that the optimal power allocation for
this case is the well-known ``water-filling'' policy
\cite{Goldsmith}. Let
$\tilde{p}=\frac{1}{\beta}-\frac{I+\sigma^2}{h}$. The optimal power
allocation for information transfer can be expressed as
\begin{align}p_{{\rm ID}}=[\tilde{p}]_0^{P_{{\rm
peak}}},\end{align}where $[x]_a^b\triangleq \max(\min(x,b),a)$.

When $\rho=0$, it follows that $L_\nu^{{\rm R-E}}(p,\rho=0)$ has the
same expression as that given in (\ref{eqn:rho=0}), and
consequently, the optimal power allocation for EH mode, $p_{{\rm
EH}}$, is given by (\ref{eqn:optimal power for EH mode}).

To summarize, for ID mode, if $\frac{1}{\beta}>P_{{\rm peak}}$, we
have
\begin{align}\label{eqn:Lagrangian ID 4 case 1} L_{\nu}^{{\rm
R-E}}(p_{{\rm
ID}},\rho=1)=\left\{\begin{array}{ll}\log(1+\frac{hP_{{\rm
peak}}}{I+\sigma^2})-\beta P_{{\rm peak}}, &
\frac{h}{I+\sigma^2}\geq \frac{1}{\frac{1}{\beta }-P_{{\rm peak}}}
\\ \log\left(\frac{h}{\beta
(I+\sigma^2)}\right)-\left(1-\frac{\beta(I+\sigma^2)}{h}\right),
& \beta  \leq \frac{h}{I+\sigma^2}<\frac{1}{\frac{1}{\beta }-P_{{\rm peak}}} \\  0. & {\rm otherwise}. \\
\end{array} \right.  \end{align}If $\frac{1}{\beta}\leq P_{{\rm peak}}$, we have \begin{align}\label{eqn:Lagrangian ID 4 case 2}L_\nu^{{\rm
R-E}}(p_{{\rm
ID}},\rho=1)=\left\{\begin{array}{ll}\log\left(\frac{h}{\beta
(I+\sigma^2)}\right)-\left(1-\frac{\beta(I+\sigma^2)}{h}\right),
& \frac{h}{I+\sigma^2}\geq \beta  \\  0, & {\rm otherwise} \\
\end{array} \right.  \end{align}For EH mode, the expression of
$L_\nu^{{\rm R-E}}(p_{{\rm EH}},\rho=0)$ is the same as that given
in (\ref{eqn:EH utility}).

Then, given a pair of $\lambda$ and $\beta$, the optimal solution to
Problem (\ref{eqn:subproblem4}) for fading state $\nu$ can be
expressed as\begin{eqnarray}&& \rho^\ast=\left\{\begin{array}{ll}1,
& {\rm if} \ L_\nu^{{\rm R-E}}(p_{{\rm ID}},\rho=1)> L_\nu^{{\rm
R-E}}(p_{{\rm EH}},\rho=0) \\
0, & {\rm otherwise}; \end{array}\right. \label{eqn:optimal solution 4}\\
&& p^\ast=\left\{\begin{array}{ll}p_{{\rm ID}}, & {\rm if} \
\rho^\ast=1 \\ p_{{\rm EH}}, & {\rm if} \ \rho^\ast=0.
\end{array} \right. \label{eqn:optimal solution 4'} \end{eqnarray}Next, to find the optimal dual
variables $\lambda^\ast$ and $\beta^\ast$ for Problem (P4),
similarly as in Section \ref{sec:Optimal Solution to Problem (P2)},
the ellipsoid method can be applied. Thus, Problem (P4) is
efficiently solved.

Next, we investigate further the optimal information/energy transfer
scheduling and power control at Tx, as well as the optimal mode switching rule at
Rx. For simplicity, we only consider the case of $I(\nu)=0$, $\forall \nu$. Since there is no interference, it can be observed from
(\ref{eqn:optimal solution 4}) and (\ref{eqn:optimal solution 4'})
that there are three possible transmission modes at Tx for the case
with CSIT: ``information transfer mode'' with water-filling power
control, ``energy transfer mode'' with peak transmit power, and
``silent mode'' with no transmission, where the first transmission
mode corresponds to ID mode at Rx and the second transmission
mode corresponds to EH mode at Rx. Similar to the analysis in
Section \ref{sec:Optimal Solution to Problem (P2)}, we can define
$\mathcal{B}_{{\rm on}}^{{\rm ID}}$,
$\mathcal{B}_{{\rm on}}^{{\rm EH}}$, and
$\mathcal{B}_{{\rm off}}$ on the
non-negative $h$-axis as the regions corresponding to the above
three modes, respectively. Let $\lambda^\ast$ and $\beta^\ast$ denote the optimal dual solutions to Problem (P4). For brevity, in the following we only present the expressions of the above regions in the case of $\frac{1}{\beta^\ast}\leq P_{{\rm peak}}$. It can be shown that in this case, $\mathcal{B}_{{\rm on}}^{{\rm ID}}=\{h:\beta^\ast\sigma^2\leq h \leq h_4\}$, $\mathcal{B}_{{\rm on}}^{{\rm EH}}=\{h:h>h_4\}$ and $\mathcal{B}_{{\rm off}}=\{h:h< \beta^\ast \sigma^2\}$, where $h_4$ is the largest root of the equation: $\log\frac{h}{\beta^\ast\sigma^2}-1+\frac{\beta^\ast\sigma^2}{h}-\lambda^\ast hP_{{\rm peak}}+\beta^\ast P_{{\rm peak}}=0$, which can be obtained by the bisection method over the interval $(\frac{\beta^\ast}{\lambda^\ast},\infty)$. The proof is omitted here due to the space limitation.

\begin{figure}
\centering
 \epsfxsize=0.8\linewidth
    \includegraphics[width=10cm]{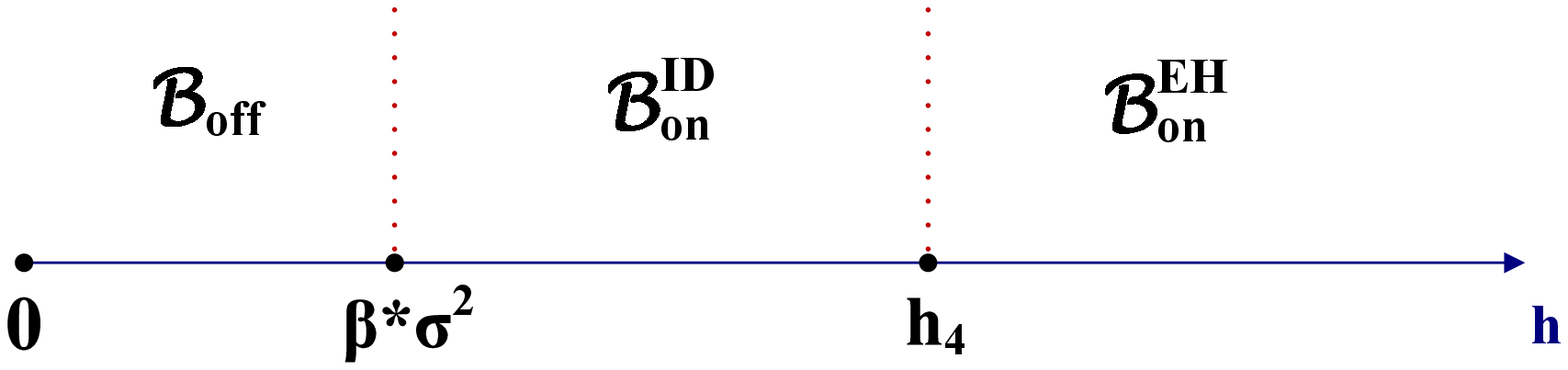}
\caption{Illustration of the optimal transmitter and receiver modes
for characterizing R-E trade-offs in the case with CSIT. It is assumed that $I(\nu)=0$, $\forall \nu$, and
$\frac{1}{\beta^\ast}<P_{{\rm peak}}$.}
\label{fig7}
\end{figure}

An illustration of $\mathcal{B}_{{\rm on}}^{{\rm ID}}$,
$\mathcal{B}_{{\rm on}}^{{\rm EH}}$, and
$\mathcal{B}_{{\rm off}}$ in the case without interference and $\beta^\ast\leq \frac{1}{P_{{\rm peak}}}$ is given in Fig. \ref{fig7}. Compared with the case without CSIT (cf. Fig. \ref{fig6}), it can be similarly observed that the channels with largest power are allocated to EH mode. However, when the channel condition is very poor, the transmitter will shut down its transmission to save power in the case with CSIT, instead of transmitting constant power in the case without CSIT.

\section{Consideration of Receiver Energy Consumption}\label{sec:The Case When the Circuit Power Consumption is Considered for Both ID Mode and EH Mode}
In the above analysis, we have ignored energy consumptions at the receiver for the purpose of exposition. In this section, we extend the result by considering the receiver energy consumption. Firstly, we explain in more details the operations of the receiver in each block and their corresponding energy consumptions as follows. At the beginning of each block, the receiver estimates the channel and interference power gains to determine which of the EH/ID mode it will switch to, where we assume a constant energy $Q_0$ being consumed. After that, suppose the receiver switches to EH mode. Since practical energy receivers are mostly passive \cite{Rui12}, we assume that the energy consumed by the energy receiver is negligibly small and thus can be ignored. However, if the receiver switches to ID mode, more substantial energy consumption is required \cite{Rui12}; for simplicity, we assume that a constant power $P_I$ incurs due to the information receiver when it is switched on. In the following, we will study the effect of the above receiver power consumptions on the optimal operation of the time-switching receiver. Due to the space limitation, we will only study the O-E trade-off in the case without CSIT, while similar results can be obtained for other cases.

Let $Q_I(\nu)=\rho(\nu)P_I$ denote the receiver power consumption due to ID mode at fading state $\nu$, and $\bar{Q}$ denote the net harvested energy obtained by subtracting $Q_0$ and $E_\nu[Q_I(\nu)]$ from the harvested energy $E_\nu[Q(\nu)]$. To study the O-E trade-off in the case without CSIT, we modify Problem (P1) as\begin{align*}\mathrm{(P5)}:~\mathop{\mathtt{Maximize}}_{\{\rho(\nu)\}}
& ~~~ E_\nu[X(\nu)] \\
\mathtt {Subject \ to} & ~~~ E_\nu[Q(\nu)]-E_\nu[Q_I(\nu)]-Q_0\geq \bar{Q} \\ & ~~~
\rho(\nu)\in \{0,1\}, \ \forall \nu
\end{align*}Since $Q_0$ is a constant for all fading states, without loss of generality we absorb this term into $\bar{Q}$ and assume $Q_0=0$ in the rest of this paper for convenience.

\begin{figure}
\centering
 \epsfxsize=0.7\linewidth
    \includegraphics[width=10cm]{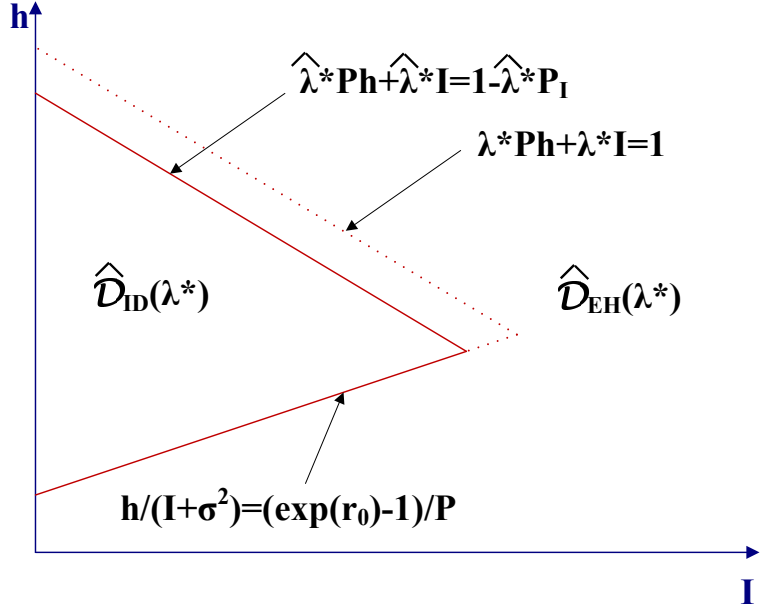}
\caption{Illustration of the optimal ID and EH regions for
characterizing O-E trade-offs with versus without receiver energy consumption in the case without CSIT.}\label{fig10}
\end{figure}

Let $\hat{\lambda}^\ast$ denote the optimal
dual variable corresponding to the net harvested energy
constraint. We then solve Problem (P5) in a similar way as for Problem (P1). The optimal solution of Problem (P5) can be expressed as\begin{align}\rho^\ast=\left\{\begin{array}{ll}1, & {\rm if} \
\frac{h}{I+\sigma^2}>\frac{e^{r_0}-1}{P} \ {\rm and} \ \hat{\lambda}^\ast
hP+\hat{\lambda}^\ast I<1-\hat{\lambda}^\ast P_I \\ 0, & {\rm otherwise}. \end{array} \right.
\end{align}
As a result, the optima ID region when the receiver energy consumption is considered can be defined as\begin{align}\label{region2}\hat{\mathcal{D}}_{{\rm
ID}}(\hat{\lambda}^\ast)\triangleq\big\{(h,I):\frac{hP}{I+\sigma^2}\geq e^{r_0}-1,1-\hat{\lambda}^\ast P_I\geq \hat{\lambda}^\ast hP+\hat{\lambda}^\ast I, \ h\geq 0, \ I\geq 0\big\},\end{align}and the rest of the plane is the optimal EH region. An illustration of the optimal ID region and EH region is given in Fig. \ref{fig10}. By comparing it with Fig. \ref{fig4} for the case without considering the receiver energy consumption, we observe that to harvest the same amount of net energy we need to allocate more fading states in (\ref{eqn:optimal ID region}) to EH mode, i.e., allocating all $(h,I)$ pairs satisfying $\frac{1}{\hat{\lambda}^\ast}-P_I\leq hP+I \leq \frac{1}{\lambda^\ast}$ to EH mode with $P_I>0$.

\begin{figure}
\centering
 \epsfxsize=0.7\linewidth
    \includegraphics[width=12cm]{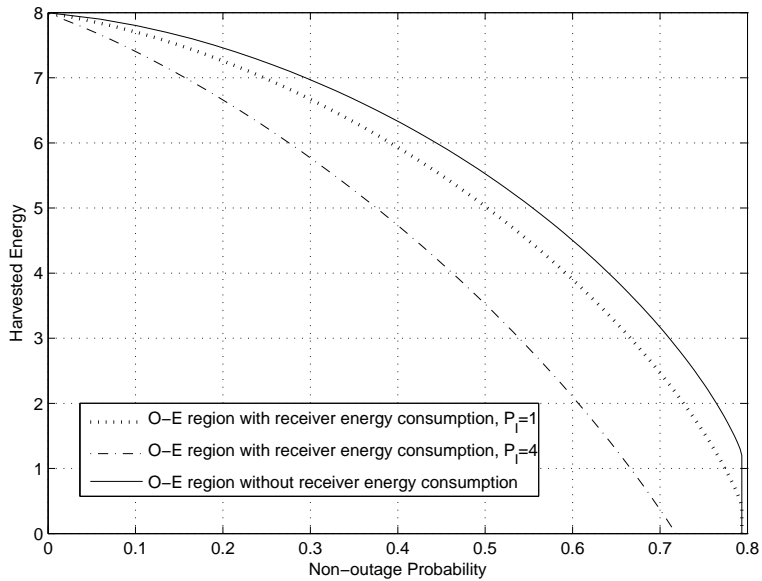}
\caption{O-E region with versus without receiver energy consumption in the case without CSIT.}
\label{fig12}
\end{figure}

Fig. \ref{fig12} shows an example of the O-E region without CSIT but considering the receiver power consumption. The setup is the same as that for Fig. \ref{fig3}. It is observed that the receiver power consumption degrades the O-E trade-off. However, $Q_{{\rm max}}$ does not change the value because it is achieved when all the fading states are allocated to EH mode and thus $P_I$ has no effects. Moreover, it is observed that when $P_I=1$, the same maximum non-outage probability $\delta_{{\rm max}}$ as that of the case without receiver energy consumption (i.e., $P_I=0$) is achieved, while when $P_I=4$, a smaller $\delta_{{\rm max}}$ is achieved. The reason is as follows. If $P_I$ is not large enough, the energy harvested in the outage fading states can offset the receiver power consumption in the non-outage fading states. As a result, all the non-outage fading states can still be allocated to ID mode. Otherwise, if $P_I$ is too large, then we have to sacrifice some non-outage fading states to EH mode to harvest more energy for ID mode, and thus the value of $\delta_{{\rm max}}$ is reduced.

\section{Numerical Results}\label{sec:simulation results}

In this section, we evaluate the performance of the proposed optimal
schemes as compared to three suboptimal schemes (to be given later)
that are designed to reduce the complexity at Rx and thus yields
suboptimal O-E or R-E trade-offs. We assume that Rx needs to have an
average harvested energy $\bar{Q}$ to maintain its normal operation.
Thus, with a given $\bar{Q}$, we will compute and then compare the
minimum outage probability or the maximum ergodic capacity
achievable by the optimal and suboptimal schemes.

First, we introduce three suboptimal receiver mode switching rules,
namely, \emph{Periodic Switching}, \emph{Interference-Based
Switching}, and \emph{SINR-Based Switching} as follows.

\begin{itemize}
\item {\bf Periodic Switching}: In this scheme, Rx switches between ID mode and EH
mode periodically regardless of the CSI. For convenience, let
$\theta$ with $0\leq \theta \leq 1$ denote the portion of time
switched to EH mode; then $1-\theta$ denotes the portion of time for
ID mode. The value of $\theta$ is determined such that the given
energy constraint $\bar{Q}$ is satisfied. For example, for the O-E
trade-off without CSIT, the maximum harvested energy $Q_{{\rm max}}$
is given in (\ref{eqn:maxenergy}). Thus, $\theta$ can be obtained as
$\theta=\frac{\bar{Q}}{Q_{{\rm max}}}$. For other trade-off cases,
$\theta$ can be obtained similarly.

\item {\bf Interference-Based Switching}: In this scheme, we assume that Rx's mode
switching is determined solely by the interference power $I(\nu)$.
When $I(\nu)>I_{{\rm thr}}$ where $I_{{\rm thr}}$ denotes a
preassigned threshold, Rx switches to EH mode; otherwise, it
switches to ID mode. The value of $I_{{\rm thr}}$ is determined so
as to meet the given energy constraint $\bar{Q}$, and the derivation
of $I_{{\rm thr}}$'s for different trade-off cases are omitted for
brevity.


\item {\bf SINR-Based Switching}: In this scheme, the mode switching is based on the
receiver's signal-to-noise-plus-interference ratio (SINR) $\frac{h(\nu)}{I(\nu)+\sigma^2}$. If
$\frac{h(\nu)}{I(\nu)+\sigma^2}>\Gamma_{{\rm thr}}$ where
$\Gamma_{{\rm thr}}$ denotes a predesigned SINR threshold, Rx
switches to ID mode; otherwise, it switches to EH mode. The value of
$\Gamma_{{\rm thr}}$ is determined so as to meet the given energy
constraint $\bar{Q}$, while the derivation of $\Gamma_{{\rm thr}}$'s
for different trade-off cases are omitted due to the space
limitation.
\end{itemize}

Moreover, if CSIT is available, Tx can implement the optimal power
control to minimize the outage probability or maximize the ergodic
capacity for information transfer, according to each of the
above three suboptimal Rx's mode switching rules.

\begin{figure}
\centering
 \epsfxsize=0.8\linewidth
    \includegraphics[width=12cm]{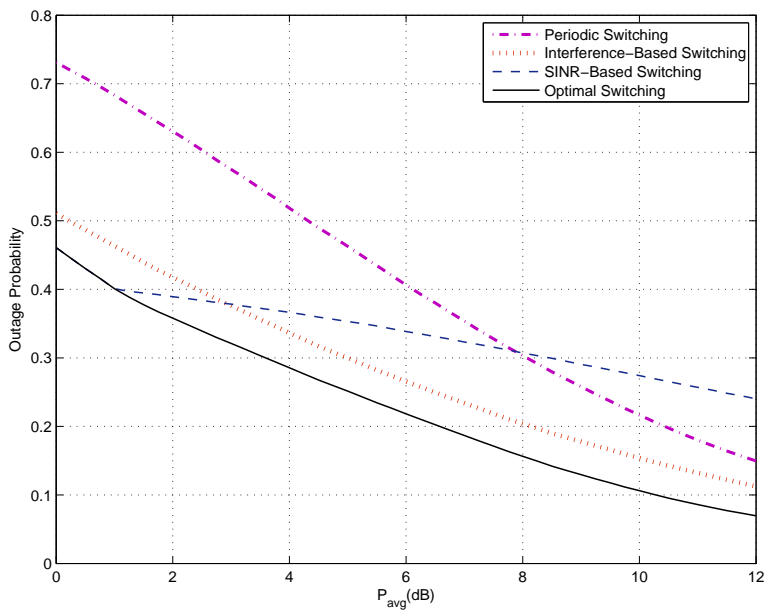}
\caption{Outage probability comparison for delay-limited information transfer in the case without CSIT and $\bar{Q}=2$.} \label{fig8}
\end{figure}

\begin{figure}
\centering
 \epsfxsize=0.8\linewidth
    \includegraphics[width=12cm]{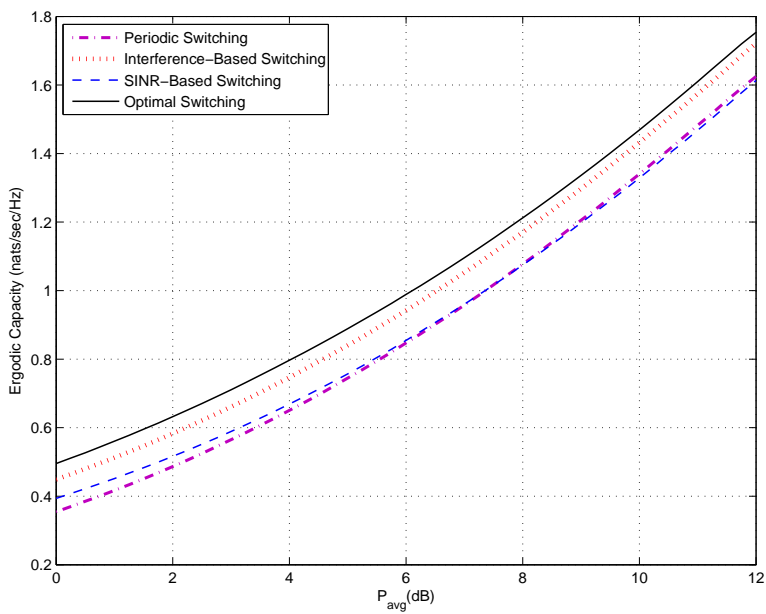}
\caption{Ergodic capacity comparison for no-delay-limited information transfer in the case with CSIT and $\bar{Q}=2$.} \label{fig9}
\end{figure}

Next, we show the performance comparison of the three suboptimal
schemes with the optimal scheme given in Section \ref{sec:Optimal
Solution to Problem (P1)} for delay-limited transmission without
CSIT and that given in Section \ref{sec:Optimal Solution to Problem
(P4)} for no-delay-limited transmission with CSIT in Figs.
\ref{fig8} and \ref{fig9}, respectively. The setup is as follows.
The PPC is $P_{{\rm peak}}=20$, the noise power is $\sigma^2=0.5$,
and for the O-E case, the constant rate requirement is
$r_0=0.2 \ {\rm nats}/{\rm sec}/{\rm Hz}$. We further
assume that $h(\nu)$ and $I(\nu)$ are independent exponentially
distributed RVs with mean $1$ and $3$, respectively. In addition, the energy target at Rx is set to be $\bar{Q}=2$.

Fig. \ref{fig8} shows the achievable minimum outage probability of
different schemes with given $\bar{Q}=2$ for the delay-limited
information transmission without CSIT. It is observed that in
general the interference-based switching works pretty well since its
performance is similar to that of the optimal switching derived in Section \ref{sec:Optimal Solution to Problem (P1)} for all values of
$P_{{\rm avg}}$ with only a small gap. On the contrary, the periodic
switching rule does not perform well with an outage probability
loss of about $10\%-20\%$ as compared to the optimal switching.

Another interesting observation is on the performance of the
SINR-based switching. It is observed from Fig. \ref{fig8} that
when $P_{{\rm avg}}\leq 1{\rm dB}$, the performance of SINR-based switching is the same as
that of the optimal switching. However, as $P_{{\rm
avg}}$ increases, its performance degrades. When $P_{{\rm
avg}}>8{\rm dB}$, its achievable outage probability is even higher
than that of periodic switching. The above observations can be
explained as follows. It can be seen from (\ref{eqn:minenergy}) in
Appendix that if we view $Q_{{\rm min}}$ as a
function of $P$, the following trade-off arises: if the value of $P$
is larger, less number of fading states are allocated to EH mode,
but more energy are harvested in each fading state allocated to EH
mode. To analyze the behavior of $Q_{{\rm min}}$ over $P$, for the
case with $h(\nu)\sim {\rm exp}(\lambda_1)$ and $I(\nu)\sim {\rm
exp}(\lambda_2)$, we can derive an explicit expression of $Q_{{\rm
min}}$ as follows:\begin{align}Q_{{\rm min}}\triangleq
f(P)=-\frac{\lambda_2e^{-\frac{\lambda_1(e^{r_0}-1)\sigma^2}{P}}P}{\lambda_2P+\lambda_1(e^{r_0}-1)}\big(\frac{e^{r_0}P}{\lambda_2P+\lambda_1(e^{r_0}-1)} +\frac{P}{\lambda_1}+(e^{r_0}-1)\sigma^2\big)+\frac{1}{\lambda_2}+\frac{P}{\lambda_1}.\end{align}
It can be shown that in our setup ($\lambda_1=1$,
$\lambda_2=\frac{1}{3}$, $r_0=0.2$ and $\sigma^2=0.5$), $f(P)$ is a
monotonically decreasing function with respect to $P$ when $0{\rm
dB}\leq P \leq 12{\rm dB}$. Moreover, when $P=1{\rm dB}$,
$f(P)=1.9998$. Thus, if $P\leq 1{\rm dB}$, it follows that $Q_{{\rm
min}}\geq \bar{Q}=2$. In other words, if $P\leq 1{\rm dB}$, the
minimum outage probability with harvested energy constraint
$\bar{Q}=2$ is achieved when Rx switches to ID mode in the fading
states
$\mathcal{H}=\{(h,I)|\log\left(1+\frac{hP}{I+\sigma^2}\right)\geq
r_0\}$ and switches to EH mode in any subset of
$\bar{\mathcal{H}}=\mathbb{R}^2_+\backslash \mathcal{H}$ to meet the
energy constraint. Consequently, the SINR-based switching
is optimal when $P$ is small. When $P>1{\rm dB}$, the minimum
harvested energy $Q_{{\rm min}}$ cannot meet the energy constraint,
and as shown in Section \ref{sec:Optimal Solution to Problem (P1)},
the optimal switching is to allocate some fading states with the
largest value of $hP+I$ in $\mathcal{H}$ to EH mode. However, the
SINR-based switching does the opposite way: it tends to allocate the
fading states with small value of $h$ to EH mode. Thus, when $P$ is
large and a certain number of fading states are allocated to EH
mode, the incremental harvested energy by the SINR-based switching
is far from that by the optimal switching. To recover this energy
loss, more fading states need to be allocated to EH mode. This is
why the SINR-based switching results in very high outage probability
when $P$ becomes large.

Fig. \ref{fig9} shows the achievable maximum rate of different
schemes with given $\bar{Q}=2$ for the no-delay-limited
information transmission with CSIT. Similar to Fig. \ref{fig8}, it
is observed from Fig. \ref{fig9} that the performance of the
interference-based switching is very close to that of the optimal switching
derived in Section \ref{sec:Optimal Solution to Problem (P4)}, while
the performances of the other two suboptimal switching rules are
notably worse. Under certain conditions (e.g., when SNR$>8$dB in Fig. \ref{fig9}), the performance of the SINR-based switching can be even worse than that of the periodic switching. This is as expected since although high SINR is preferred by information decoding, the optimal mode switching rule derived in Section \ref{sec:Optimal Solution to Problem (P4)} is determined by both the values of $h$ and $I$, but has no direct relationship to the ratio of them, i.e., the SINR value. Thus, the performance of the SINR-based switching cannot be guaranteed.

\section{Concluding Remarks}\label{sec:concluding remarks}
This paper studied an emerging application in wireless communication
where the receiver opportunistically harvests the energy from the
unintended interference and/or intended signal in addition
to decoding the information. Under a point-to-point flat-fading
channel setup with time-varying interference, we derived the
optimal ID/EH mode switching rules at the receiver to optimize the
outage probability/ergodic capacity versus harvested energy
trade-offs. When the CSI is known at the transmitter, joint
optimization of transmitter information/energy scheduling and power
control with the receiver ID/EH mode switching was also
investigated. Somehow counter-intuitively, we showed that for
wireless information transfer with opportunistic energy harvesting,
the best strategy to achieve the optimal O-E and R-E trade-offs is
to allocate the fading states with the best direct channel gains to
power transfer rather than information transfer. Moreover, three
heuristic mode switching rules were proposed to reduce the
complexity at Rx, and their performances were compared against the
optimal performance.

There are important problems unaddressed yet in this paper and thus
worth further investigation, some of which are
highlighted as follows:
\begin{itemize}

\item In this paper, we assumed that the interference is within the same band as the transmitted signal from
Tx. As a result, the algorithms proposed in this paper to achieve
the optimal O-E or R-E trade-offs cannot be directly applied to the case of
wide-band interference. It is thus interesting to investigate how to
manage the wide-band interference in a wireless energy harvesting
communication system.

\item In this paper, we studied the optimal mode switching and/or power control rules in a single-user setup subject to
an aggregate interference at the receiver. However, how to extend the results of this paper to the multi-user setup is an unsolved problem.
For the multi-user interference channel, interference management is a key issue. Traditionally,
interference is either decoded and subtracted when it is strong or treated as noise when it is weak.
In this paper, we provide a new approach to deal with the interference by utilizing it as a new source for energy harvesting.
Thus, how should the Tx-Rx links in an interference channel cooperate with each other to manage the interference by
optimally balancing between information and power transfer is an intricate problem requiring further investigation.

\end{itemize}

\begin{appendix}

In this appendix, we characterize the vertex points on the
boundary of the O-E region and R-E region (cf. Fig. \ref{fig3}) for both the cases with
and without CSIT.

\subsubsection{O-E region without CSIT}

\ \ \

As shown in Fig. \ref{fig3}(a), $Q_{{\rm max}}$ is given
by\begin{align}\label{eqn:maxenergy}Q_{{\rm
max}}=E_{\nu}[h(\nu)P+I(\nu)],\end{align}when $\rho(\nu)=0$,
$\forall \nu$, i.e., EH mode is active all the time at Rx and thus
the resulting non-outage probability $\delta_{{\rm min}}=0$
(corresponding to the outage probability equal to 1). Moreover,
$Q_{{\rm min}}$ and $\delta_{{\rm max}}$ are given by\begin{align} &
Q_{{\rm
min}}=\int\limits_{\nu:\log\big(1+\frac{h(\nu)P}{I(\nu)+\sigma^2}\big)<r_0}\big(h(\nu)P+I(\nu)\big)f_\nu(h,I)d\nu,\label{eqn:minenergy}\\
& \delta_{{\rm
max}}=Pr\left\{\log\left(1+\frac{h(\nu)P}{I(\nu)+\sigma^2}\right)\geq
r_0\right\}.
\end{align} Note that $Q_{{\rm min}}$ is the minimum average harvested energy at Rx when the maximum non-outage probability (or minimum outage probability) is achieved. Since the set for the outage fading states is non-empty in (\ref{eqn:minenergy}), $Q_{{\rm min}}\neq 0$ in general.

\subsubsection{O-E region with CSIT}

\ \ \

As shown in Fig. \ref{fig3}(a), the point $(\delta_{{\rm
min}},Q_{{\rm max}})$ is achieved when all the fading states are
allocated to EH mode, i.e., $\rho(\nu)=0$, $\forall \nu$. Thus, the
resulting non-outage probability is $\delta_{{\rm min}}=0$. Moreover,
the harvested energy can be expressed as
$Q=E_\nu[h(\nu)p(\nu)]+E_\nu[I(\nu)]$, where the first term is the
energy harvested from the signal, while the second term is due to
the interference. To maximize the first term under both the PPC and
APC, the optimal power control policy is to transmit at peak power
at the fading states with the largest possible $h$'s. Let
$\hat{h}_1$ be the threshold that
satisfies\begin{align}\underset{\nu:h(\nu)\geq\hat{h}_1}{\int}
P_{{\rm peak}}f_\nu(h,I)d\nu=P_{{\rm avg}}.\end{align}Then $Q_{{\rm
max}}$ can be expressed
as\begin{align}\label{eqn:maxenergycsi}Q_{{\rm
max}}=\underset{\nu:h(\nu)\geq\hat{h}_1}{\int} h(\nu)P_{{\rm
peak}}f_\nu(h,I)d\nu+E_\nu[I(\nu)].\end{align}

To obtain $\delta_{{\rm max}}$, we need to minimize the outage
probability under both the APC and PPC without presence of the
energy harvester. It can be shown that the optimal power allocation
to achieve the maximum non-outage probability can be expressed as
the well-known truncated channel inversion policy \cite{Caire},
\cite{Goldsmith}:\begin{align}p^\ast(\nu)=\left\{\begin{array}{ll}\frac{(e^{r_0}-1)(I(\nu)+\sigma^2)}{h(\nu)},
& {\rm if} \ \frac{h(\nu)}{I(\nu)+\sigma^2}\geq \hat{h}_2. \\ 0, &
{\rm otherwise} \end{array}\right. \end{align}where
$\hat{h}_2=\max\{\beta(e^{r_0}-1),\frac{e^{r_0}-1}{P_{{\rm
peak}}}\}$ with $\beta$ denoting the optimal dual variable
associated with the APC that satisfies $E_\nu[p^\ast(\nu)]=P_{{\rm
avg}}$. Then the maximum non-outage can be expressed
as\begin{align}\delta_{{\rm
max}}=Pr\left\{\frac{h(\nu)}{I(\nu)+\sigma^2}\geq
\hat{h}_2\right\}.\end{align}On the other hand, $Q_{{\rm min}}$ is
achieved when Rx harvests energy at all the outage fading states.
Let $\hat{h}_3$ denote the value of $h$ that
satisfies\begin{align}\underset{\nu:h(\nu)\geq\hat{h}_3,\frac{h(\nu)}{I(\nu)+\sigma^2}\leq\hat{h}_2}{\int}P_{{\rm
peak}}f_\nu(h,I)d\nu+\underset{\frac{h(\nu)}{I(\nu)+\sigma^2}\geq\hat{h}_2}{\int}p^\ast(\nu)f_\nu(h,I)d\nu=P_{{\rm
avg}}.\end{align}Then the minimum harvested energy can be expressed
as\begin{align}Q_{{\rm
min}}= \underset{\nu:h(\nu)\geq\hat{h}_3,\frac{h(\nu)}{I(\nu) +\sigma^2}\leq\hat{h}_2}{\int}hP_{{\rm
peak}}f_\nu(h,I)d\nu+\underset{\nu:\frac{h(\nu)}{I(\nu)+\sigma^2}\leq\hat{h}_2}{\int}I(\nu)f_\nu(h,I)d\nu.\end{align}Note
that if
$\underset{\frac{h(\nu)}{I(\nu)+\sigma^2}\geq\hat{h}_2}{\int}p^\ast(\nu)f_\nu(h,I)d\nu\geq
P_{{\rm avg}}$, then $\hat{h}_3=\infty$, i.e., no power is available
for energy transfer at Tx. Thus, $Q_{{\rm min}}$ is only due to the
interference power. Since the set for the outage fading states is non-empty, $Q_{{\rm min}}\neq 0$ since the receiver can at least harvest energy from the interference in the outage fading states.

\subsubsection{R-E region without CSIT}

\ \ \

As shown in Fig. \ref{fig3}(b), the maximum harvested energy
$Q_{{\rm max}}$ is achieved when all the fading states are allocated
to EH mode, i.e., $\rho(\nu)=0$, $\forall \nu$, and thus has the
same expression as that given in (\ref{eqn:maxenergy}). Moreover,
$R_{{\rm min}}=0$. On the other hand, the ergodic capacity is
maximized when all the fading states are allocated to ID mode, i.e.,
$\rho(\nu)=1$, $\forall \nu$. Consequently, $Q_{{\rm min}}=0$ and
\begin{align}R_{{\rm
max}}=E_\nu\left[\log\left(1+\frac{h(\nu)P}{I(\nu)+\sigma^2}\right)\right].\end{align}

\subsubsection{R-E region with CSIT}

\ \ \

As shown in Fig. \ref{fig3}(b), similar to the case of O-E region
with CSIT, the maximum harvested energy $Q_{{\rm max}}$ is given in
(\ref{eqn:maxenergycsi}), and $R_{{\rm min}}=0$. As for the point
$(R_{{\rm max}},Q_{{\rm min}})$, to maximize the ergodic capacity
under both the APC and PPC, the optimal transmit power policy is the
well-known ``water-filling`` power allocation given by
\cite{Goldsmith}
\begin{align}p^\ast(\nu)=\left[\frac{1}{\lambda^\ast}-\frac{I(\nu)+\sigma^2}{h(\nu)}\right]_0^{P_{{\rm
peak}}},\end{align}where $[x]_a^b\triangleq \max(\min(x,b),a)$, and
$\lambda^\ast$ is the optimal dual variable associated with $P_{{\rm
avg}}$ satisfying $E_\nu[p^\ast(\nu)]=P_{{\rm avg}}$. Thus, the
maximum rate is given by\begin{align}R_{{\rm
max}}=E_\nu\left[\log\left(1+\frac{h(\nu)p^\ast(\nu)}{I(\nu)+\sigma^2}\right)\right].\end{align}Then,
for the fading states satisfying
$\frac{h(\nu)}{I(\nu)+\sigma^2}<\lambda^\ast$, Rx can harvest energy
from the interference. Thus the minimum harvested energy is in general non-zero and can be
expressed as\begin{align}Q_{{\rm
min}}=\underset{\nu:\frac{h(\nu)}{I(\nu)+\sigma^2}<\lambda^\ast}{\int}I(\nu)f_\nu(h,I)d\nu.\end{align}

\end{appendix}

\end{document}